\documentclass[twocolumn,aps,showpacs,amssymb]{revtex4}
\input epsfig.sty
\newcommand{\be}{\begin{equation}}
\newcommand{\ee}{\end{equation}}
\newcommand{\beq}{\begin{eqnarray}}
\newcommand{\eeq}{\end{eqnarray}}
\usepackage{bm}
\usepackage{graphicx}%

\begin{document}
\setcounter{figure}{\arabic{figure}}

\title{ The Matter and the Pseudoscalar Densities in Lattice QCD} 
\author{C.~Alexandrou}
\affiliation{ Department of Physics, University of Cyprus, P.O. Box 20537,
CY-1678 Nicosia, Cyprus }
\author{Ph.~de~Forcrand} 
\affiliation{Institute for Theoretical Physics, ETH H\"onggerberg,
 CH-8093 Z\"urich, Switzerland\\ and
CERN, Theory Division, CH-1211 Geneva 23, Switzerland}
\author{A.~Tsapalis} 
\affiliation { Department of Physics, University of Cyprus, P.O. Box 20537,
CY-1678 Nicosia, Cyprus} 

\date{\today}%

\begin{abstract}  
The matter and the pseudoscalar densities
 inside a hadron are calculated via
 gauge-invariant equal-time correlation functions. A comparison is
made between the  charge and the matter density 
distributions for the pion, the rho,
the nucleon and the $\Delta^+$ within the quenched theory, and   
with two flavours of  dynamical quarks. 
\end{abstract}


\maketitle 

 
\section{Introduction}
The one-body matter density distribution plays  an important role
in our understanding of nuclear structure. In
nuclei  it determines the intranuclear 
distribution which includes both protons and neutrons,
unlike the charge  density which determines the proton distribution.
It requires hadronic probes and for this reason it is  not as accurately
determined as the charge distribution. However at new
experimental facilities such as RIKEN in Japan there are plans to 
improve the  measurements of  the matter density distribution 
using proton beams~\cite{japan}. Similarly matter density distributions of 
individual hadrons
are determined 
 in  hadron proton collisions ~\cite{Akerlof}.  From
the analysis of elastic differential
cross sections for $\pi^{\pm} p$ and $K^{\pm} p$, the radii of these
mesons
can be  extracted within certain assumptions like the 
eikonal approximation and the exponentiation of the S matrix~\cite{Chou}.

In this work we study hadron matter densities using gauge invariant 
correlators calculated in lattice QCD. This is an extension of our previous
work~\cite{AFT} on hadron charge density distributions. 
In addition to the matter density distribution we present results
for the pseudoscalar density  and compare to bag model predictions.
Both quantities are evaluated in  quenched QCD and with two flavours of
dynamical quarks. 
To test our lattice procedure we perform a more detailed analysis
of the relevant
correlation functions as compared to what was done in ref.~\cite{AFT}
 extending it also to the 
matter and pseudoscalar
densities. 
By comparing the matter and charge density distributions one can draw important
phenomenological conclusions regarding hadron deformation.

\section{Gauge invariant correlation functions}
We consider the
equal-time correlators~\cite{latt,panic},

\vspace{-0.3cm}

\be
 C_\Gamma^H({\bf r},t) = \int\> d^3r'\>
\langle H|\hat{j}_\Gamma^u({\bf r'},t)\hat{j}_\Gamma^{d}({\bf r}'+{\bf r},t)|H\rangle \quad,
\ee

\vspace{-0.3cm}

\noindent
with the  current 
$\hat{j}^u_\Gamma({\bf r},t)$ given by the normal order product
$:\bar{u}({\bf r},t)\Gamma u({\bf r},t):$.
For $\Gamma =\gamma_0$ and
 $\Gamma ={\bf 1}$ we obtain
the charge and matter density distributions
respectively. 
The pseudoscalar density  obtained with   $\Gamma=\gamma_5$ 
is also evaluated since it provides a useful observable
for testing the predictions of the bag model.

\begin{figure}[h]
\begin{center}
\vspace*{0.75cm}
\epsfxsize=7.0truecm
\epsfysize=8.truecm
\mbox{\epsfbox{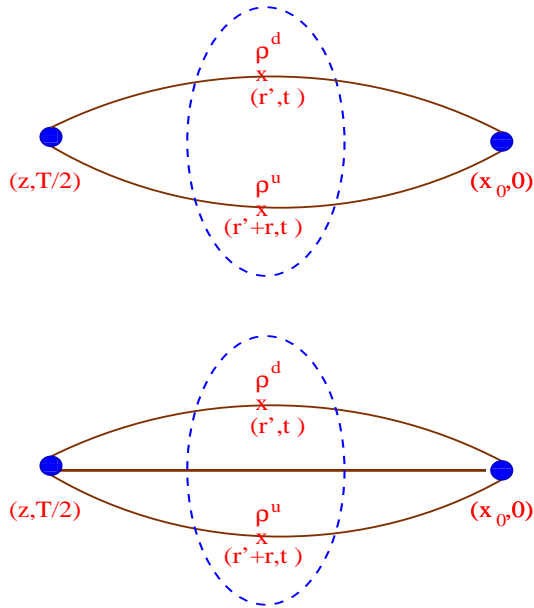}}
\caption{\label{fig:2density}Equal-time current-current correlator for a meson (upper)
and a baryon (lower). $t$ and
$T/2-t$ must be
large enough to filter out the hadronic state. $T$ is the time
extension of our lattice. We have assumed anti-periodic
boundary conditions in the temporal direction in drawing this figure.}
\end{center}
\end{figure}

The  matrix elements for mesons and baryons considered in this work 
are shown schematically in Fig.~\ref{fig:2density}.
In the case of the  baryons two relative distances are
involved and three current insertions
are required. However we may consider integrating over one relative distance
to obtain the one-particle density distribution shown schematically 
in the lower part
of Fig.~\ref{fig:2density}.
Baryon matrix elements with three  current insertions,  
one on each of the three
 quark lines, were computed in the
case of the charge density distribution
 in ref.~\cite{AFT} and shown, after integration
over one relative distance, to reproduce, within statistics, the
one-particle density.
In this work  we
will only consider  one-particle distributions,
and therefore only diagrams with two current insertions such as shown in
 Fig.~\ref{fig:2density} are evaluated.
In the future, we plan to extend our work to three current
insertions, at equal and unequal times.

We also address here an important technical point.
On our lattice of time-extent $T$, with the usual anti-periodic
boundary conditions for the fermion fields,
we insert the currents at Euclidean time separation $T/4$ 
from the source and from the sink, 
as indicated in Fig.~\ref{fig:2density}. 
The current insertions must be separated far enough from the source and the sink
to  suppress excited hadronic states and non-zero momenta.
The latter are not projected out, as often done in other studies,
because zero-momentum projection requires a summation over spatial translations
of the source or sink. Here, this summation is technically not feasible because
it would involve quark propagators from all to all spatial lattice sites.
In the following section we will demonstrate
 that our time separation of $T/4$ is sufficient to satisfactorily
filter out undesired states.

\section{Lattice techniques}

All the results presented in this work have been obtained on a lattice of size
$16^3 \times 32$. For the quenched case we use 220 NERSC~\cite{NERSC}
configurations generated at $\beta=6.0$, and for
the unquenched case we analyse two $\kappa$ values
using for each  100 SESAM configurations~\cite{SESAM} 
 simulated at 
$\beta=5.6$ with two degenerate quark flavours.
The physical volume of the lattice in the quenched and in the unquenched 
case is approximately the same. 

\begin{figure}[h]
\begin{center}
\epsfxsize=8.0truecm
\epsfysize=10.truecm
\mbox{\epsfbox{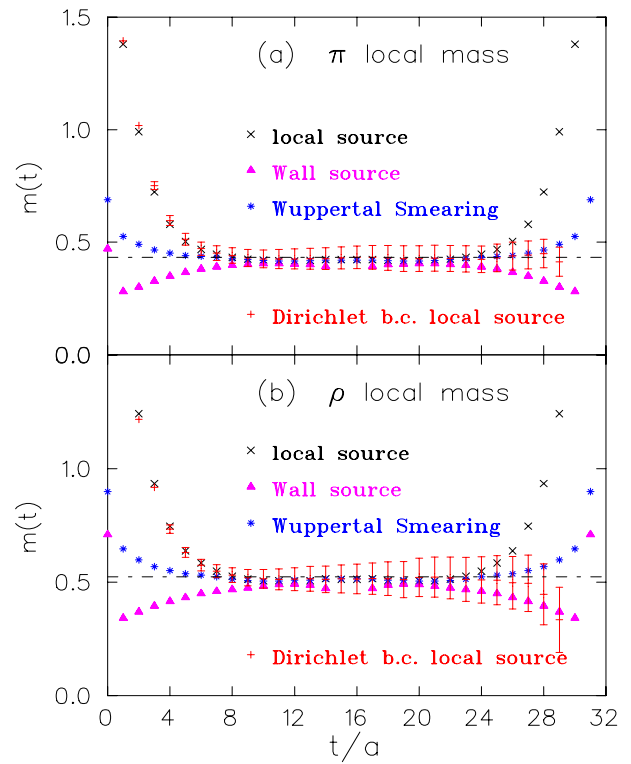}}
\caption{\label{fig:meff} The local mass, $m(t)$, 
at $\kappa=0.153$ (a) for the pion  and (b) for the rho
is shown versus the time separation from the source  in lattice units, for 
 a local source with anti-periodic boundary conditions (x), a local source with
 Dirichlet boundary conditions (crosses),
 a smeared source  (stars) and  a wall source (filled triangle).
The dashed line shows the plateau value obtained  by fitting  the data
 coming from Dirichlet boundary conditions.
}
\end{center}
\end{figure}

Let us first demonstrate that the time extent of our lattice 
is large enough to isolate the hadron ground state of zero momentum.
For this study we analysed 56 quenched  configurations at $\kappa=0.153$.  
The first check is to evaluate the mass of the particles
via the two point correlator, $C(t)$, using different sources.
For this comparison we use: i) a local source with
anti-periodic boundary conditions in the temporal direction, 
ii) a local source with
 Dirichlet boundary conditions in the temporal direction, 
iii) Wuppertal
smearing of the source and iv) a wall source which projects 
out the zero momentum {\em quark} propagator. In the latter case
we  Coulomb gauge fix 
the configuration  at the time slice of the source.
Employing Dirichlet boundary conditions allow us to utilize
the whole time extent of our lattice instead of using half as is done
for anti-periodic boundary conditions in the rest of the cases.
When using Dirichlet boundary conditions the source is placed at the second time slice,
the sink at the last but one and 
the current insertions at various intermediate
 time separations, $t$, from the source.
In all cases we use a local sink. 
The results for the  effective mass, $m(t)$, for the pion and the rho
are shown in Fig.~\ref{fig:meff}.
As expected Wuppertal smearing of the source produces the earliest plateau. 
Local sources reach the same plateau value for 
 time separations $t \ge 8a$  from the source, where $a$ is the lattice
spacing.
Using a  wall source does not improve projection to
the ground state. In fact the  effective mass still 
deviates at $t=8a$ and only
converges to the plateau value at time separations $t \ge 10a$.
For testing the mass plateaus using two point hadron correlators,
we have summed over the sink spatial volume which projects to
 the zero momentum state in the standard way. Had we not performed
the sum over the sink spatial volume the errors on the 
effective mass would have be much larger
and  the comparison of different sources would have been less meaningful.
As we have already pointed out summing over the spatial volume of
the  source is not possible
for the density distributions
 since it
would require the all-to-all propagator. 
Therefore it is crucial to check that higher momenta are sufficiently
suppressed for the evaluation of the density distributions.

\begin{figure}[h]
\begin{center}
\epsfxsize=8.0truecm
\epsfysize=10.5truecm
\mbox{\epsfbox{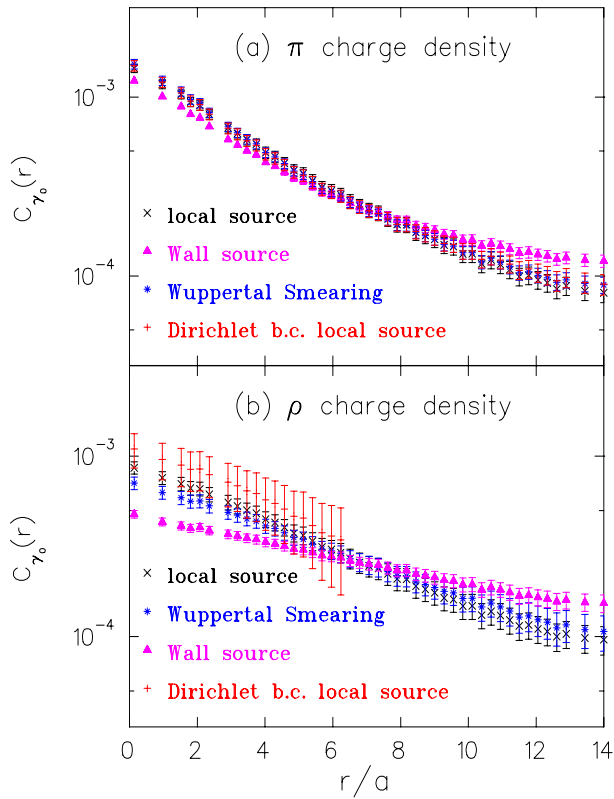}}
\caption{\label{fig:wfs_it} The charge density distribution 
at $\kappa=0.153$ (a) for the pion
and (b) for the rho is shown versus the radius in lattice units
when the current couples to the quark at time separation from the source
 of  $t=8a$ with anti-periodic boundary conditions 
and of $t=14a$ for Dirichlet boundary conditions.
In the case of the rho distribution the errors on the data obtained
with  Dirichlet boundary conditions are omitted  for distances
greater than $r/a=6$  for clarity.
The notation is as in Fig.~\ref{fig:meff}}.
\end{center}
\end{figure}

\begin{figure}[h]
\begin{center}
\epsfxsize=8.0truecm
\epsfysize=10.truecm
\mbox{\epsfbox{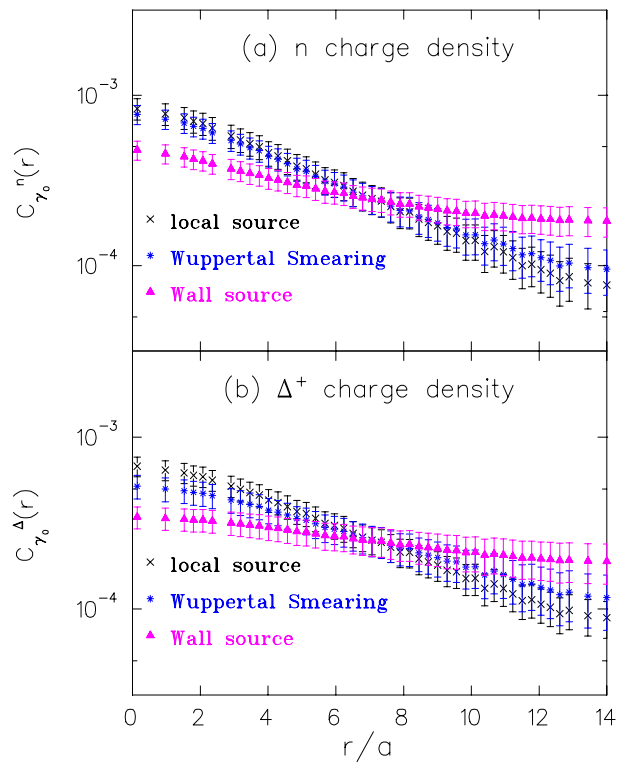}}
\caption{\label{fig:comp_source2}
The  charge density distributions at $\kappa=0.153$  versus the radius in lattice units
(a) for  the nucleon and (b) for the $\Delta^+$ using
a local source and sink with anti-periodic boundary conditions (x's), 
for a
smeared source and sink  (stars) and for a wall source and sink (filled triangle).
Data obtained using Dirichlet boundary conditions are too noisy and
they are not shown.} 
\end{center}
\end{figure} 

In  Fig.~\ref{fig:wfs_it} 
we show the results for the charge density distribution 
at different time slices
for the pion and the rho,
obtained using Dirichlet boundary conditions which allow us to
probe larger time separations. 
Here, in contrast to the two
point correlator, $C(t)$,
 the source and the sink are treated in the same way. 
For all the plots showing the
charge density distribution
 as a function of $r$ we have averaged the lattice results over
bins of size $0.28a$.
As it can be seen the results  for the pion 
 are the same at quarter lattice time 
 separation $t=8a$ and at $t=14a$.  We also have agreement between 
the results obtained with local and smeared sources.
For the rho the comparison is more difficult since
the results using Dirichlet boundary conditions are  more noisy 
than for the pion.
Within our statistics the charge density
distributions obtained at time separation $t=8a$ and $t=14a$
are the same. Also, apart from small deviations at short distances,
there is agreement of the results obtained using Wuppertal smeared and
local sources.
This means that any contamination from high momentum states
is small
and, within our statistical uncertainties, it is permissible
to  use anti-periodic
boundary conditions with current insertions at $t=8a$. 
This is fortunate, since it allows us to analyze standard full QCD
configurations.
It is also advantageous since the gauge noise is far less for time
separations $8a$.
 Using a wall source and sink
produces a correlator which is  in  disagreement with the rest
at time separation $t=8a$. This means that there is still
a sizable contamination from excited states, which is not surprising given
that the effective mass for this source has not completely converged
at $t=8a$.  
In contrast to  the other sources, where
the individual 
quark propagators can carry non-zero momentum,
 wall sources project
on zero momentum for each quark
propagator. If the time evolution is too short to filter the ground state
then the quarks  carry lower momentum than in
the exact ground state. This would explain 
 the fact that, for a wall source, the charge density distribution
is still broader than the rest at $t=8a$:   
the quarks carrying a smaller 
momentum lead to a slower decay.
This effect is less visible in the pion, which indicates that
in the ground state the quark relative momentum  is smaller than in the rho.
In
Fig.~\ref{fig:comp_source2} we compare the results obtained with the different
sources for the nucleon and the $\Delta^+$ at time separation $t=8a$.
Data with Dirichlet boundary conditions are not shown since they are too noisy.
Again
 using a wall source and sink
produces a correlator that decays more slowly than the rest, showing that 
projection to the ground state is less effective when the initial
and final states
are constructed with zero momentum quarks. On the other hand, 
there is  agreement between results obtained
with smeared and local sources
apart from
 small deviations at short distances for the  $\Delta^+$ channel. 
The independence
of the results from the different interpolating
fields indicates that the  ground state has been isolated sufficiently well. 
Therefore for the parameters used here, it is justified to use anti-periodic 
boundary conditions and local sources to  study  
the density distributions. Using Dirichlet boundary conditions
to increase the time separation between the source and the sink produces
results that
are more noisy and can not be distinguished from the ones obtained
with anti-periodic boundary conditions.

\begin{figure}[h]
\begin{center}
\epsfxsize=8.0truecm
\epsfysize=10.truecm
\mbox{\epsfbox{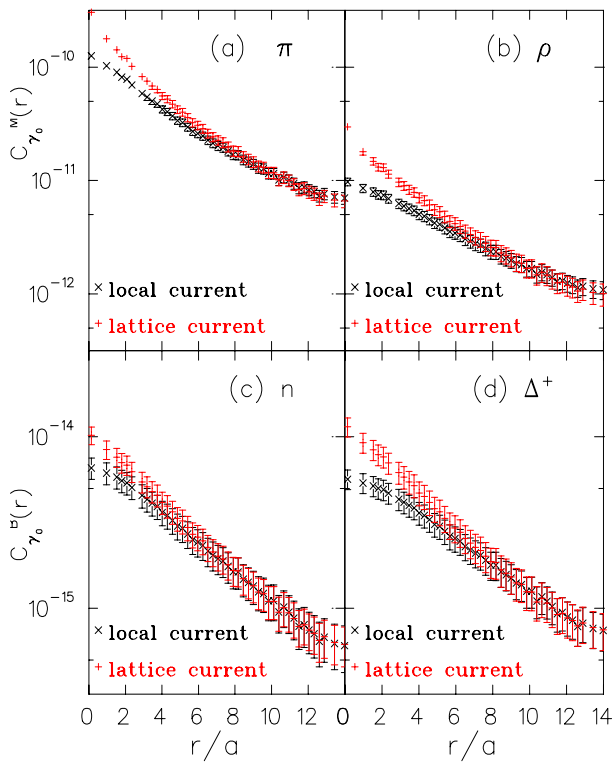}}
\caption{\label{fig:comp j}
(a) The pion, (b) the rho, (c) the nucleon and (d) 
$\Delta^+$ charge density distribution  versus the radius in lattice units
using the
local current  (x's) and the lattice conserved current (crosses). 
}
\end{center}
\end{figure}

In this work we use lattices having similar lattice spacing $a$.
In order to assess
finite lattice spacing effects on these distributions we
compare results using the continuum current
and the lattice conserved current which  differ by ${\cal O}(a)$ terms.
The lattice conserved current is given by
\beq
j_\mu (x) &=& \sum_f Q_{f} \kappa_{f} \lbrace
\bar{\psi}^{f} (x + \hat{\mu})(1 + \gamma_\mu)
U^{\mu \dagger} (x) \psi^{f} (x) \nonumber \\
&\>& -\bar{\psi}^{f} (x)(1 - \gamma_\mu)
U^{\mu} (x) \psi^{f} (x + \hat{\mu}) \rbrace 
\eeq
and it is  symmetrized at lattice site $x$. $Q_f$ is the charge of  quark
of flavor $f$. 
 The results for the unnormalized density correlator, 
shown in  Fig.~\ref{fig:comp j}, demonstrate
that finite-$a$ effects lead to the wrong behaviour near the 
origin. 
The same conclusion was also reached in the study of the density
distribution of charmonium
where  results using the Wilson Dirac operator were compared to those
using  clover and
tadpole improved operators~\cite{margarita}.
In the rest of this work, unless otherwise stated, we will normalize
the density correlators over the spatial volume
and not over their value at the origin because of the
potentially large finite $a$ error  at the origin.
An improvement beyond  the present work will be to use  an ${\cal O}(a)$-improved Dirac
operator to eliminate finite-$a$ effects at short distances.

\section{Matter density}
In this section we present results on the matter density distribution
using a local source and a local sink.
As discussed in the previous section, our best compromise for the study 
of the density distribution is to 
  fix the source and the sink for maximum separation
at $t_i=a$ and $t_f=17a$ and insert  $\bar{u}u$ and 
$\bar{d}d$ operators
half-way between at   $t-t_i=8a$. 
By
examining the dependence of the matter correlator on the time separation
$t-t_i$ at which 
the $\bar{u}u$ and 
$\bar{d}d$ operators are inserted, as done above for
the charge density,
we again conclude   that
a time interval of $|t - t_i| = |t - t_f| = 8a$ 
 is sufficient to project to the ground
state of the hadrons of interest. 
In this evaluation we use the same interpolating fields,
described in detail in ref.~\cite{AFT},
as for the density correlator, and  the Wilson Dirac operator with 
hopping parameter $\kappa=0.15,\> 0.153,\> 0.154$ and $ 0.155$.
The ratio of the pion mass to the rho mass at these values of $\kappa$  is
$0.88,\>0.84,\>0.78$ and 0.70 respectively.
Using the standard definition of the
 naive quark mass,
$2m_q a= (1/\kappa-1/\kappa_c)$, 
where $\kappa_c=0.1571$
 is the value of $\kappa$ at which
 the pion becomes massless, we find $m_q \sim 300, 170 ,130$ and $85$~MeV
respectively.
To obtain these values we  used the string tension~\cite{Bali}
to set the physical scale, which gives for the inverse lattice
spacing $a^{-1}\sim 1.94$~GeV($a=0.103$ fm). 
This choice makes direct contact to our previous study of charge
density distributions~\cite{AFT}.
Another possibility is to use the rho (or the nucleon)
 mass in the chiral limit.
Using the rho mass  yields $a^{-1}$=2.3~GeV 
($a=0.087$ fm)~\cite{QCDPAX,Gupta}, with a systematic error 
coming from the choice of fitting range and chiral extrapolation ansatz
of about 10\%,
which is about twice as large as the statistical error.
We note that Ref.~\cite{QCDPAX} observed stability of that scale with
respect to an increase of the lattice size. 
In our discussion of
quenched data we will use the value of $a$ determined from the string tension.
However, to compare the quenched  with the unquenched results we will
use the value extracted from the rho mass in the chiral limit, since this
determination is applicable both in the quenched and in the unquenched theory.
In our discussion of the bag model results we will set
the scale  using the nucleon mass since, in that case,
the nucleon mass is used to fix the bag model parameters.

\begin{figure}[h]
\begin{center}
\epsfxsize=8.0truecm
\epsfysize=10.truecm
\mbox{\epsfbox{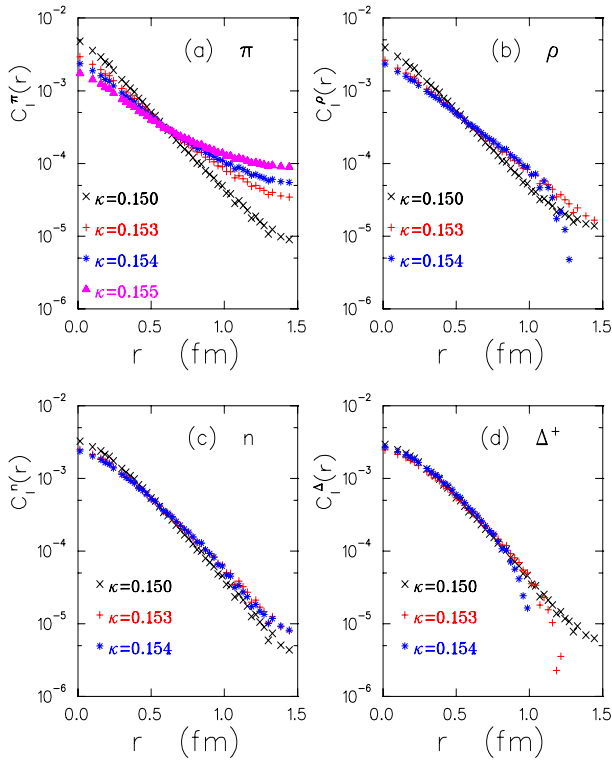}}
\caption{\label{fig:wfs matter}
(a) The pion, (b) the rho, (c) the nucleon and (d) 
$\Delta^+$ matter density distribution versus the radius for
$\kappa=0.15$ (x), $\kappa=0.153$ (crosses) and $\kappa=0.154$
(stars). For the pion we also include the
results for $\kappa=0.155$ (filled triangles). 
The errors bars are not shown for clarity.
}
\end{center}
\end{figure}

\begin{figure}[h]
\begin{center}
\epsfxsize=8.0truecm
\epsfysize=7.truecm
\mbox{\epsfbox{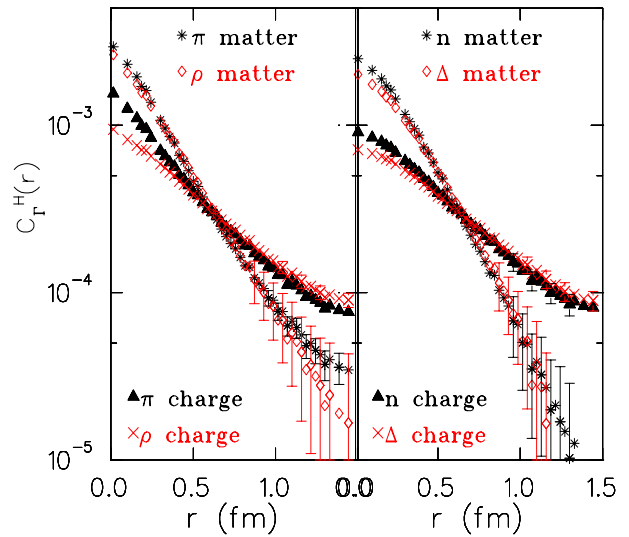}}
\caption{\label{fig:charge matter}
Comparison of the charge and matter
density distributions at $\kappa=0.153$. Left for
 the pion (filled triangles for the charge) and 
(stars  for the matter) and the rho (x for the
charge) and (open rhombus for the matter) distributions.
Right for the nucleon  (filled triangles for the charge) and 
(stars  for the matter) and the $\Delta^+$ (x for the
charge) and (open rhombus for the matter) distributions.
}
\end{center}
\end{figure}

In Fig.~\ref{fig:wfs matter} we consider the quark mass dependence of the matter
distribution.
The pion
shows the strongest dependence on the quark mass, in contrast to the charge
distribution where the rho showed the largest variation~\cite{AFT}.
For quark masses in the range of 300-100~MeV investigated here 
baryon charge and matter distributions
show essentially no variation with the quark mass.
However, charge and matter distributions are quite different from each other.
The comparison between both is shown in Fig.~\ref{fig:charge matter}
for $\kappa=0.153$. We observe, between
 the pion and the rho but not between the nucleon and the $\Delta^+$, a
larger variation for the charge density as compared to the matter
density distribution. In fact the matter density distribution
is very similar for the four hadrons considered here.
In all cases the matter distribution decays
faster than the charge density, an observation consistent with
the results of Ref.~\cite{Green}.

\begin{figure}[h]
\begin{center}
\epsfxsize=8.0truecm
\epsfysize=7.truecm
\mbox{\epsfbox{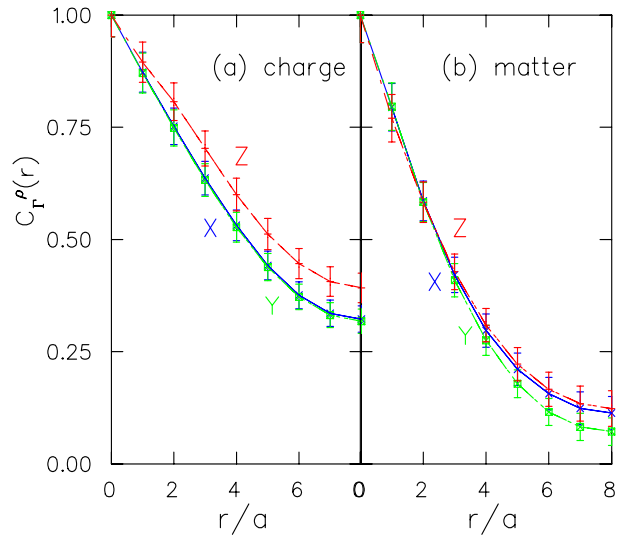}}
\caption{\label{fig:rho3 asym}
Rho asymmetry at $\kappa=0.154$ (a) for the charge and (b) for the
 matter density distributions. 
The upper curve labeled by $Z$ is $C^\rho_\Gamma(0,0,z)$
and the lower curves labeled $X$ and $Y$ is $C^\rho_\Gamma(x,0,0)$ and
 $C^\rho_\Gamma(0,y,0)$ respectively. The z-axis is along the spin direction
of the rho meson.}
\end{center}
\end{figure} 

\begin{figure}
\epsfxsize=8.truecm
\epsfysize=10.truecm
\mbox{\epsfbox{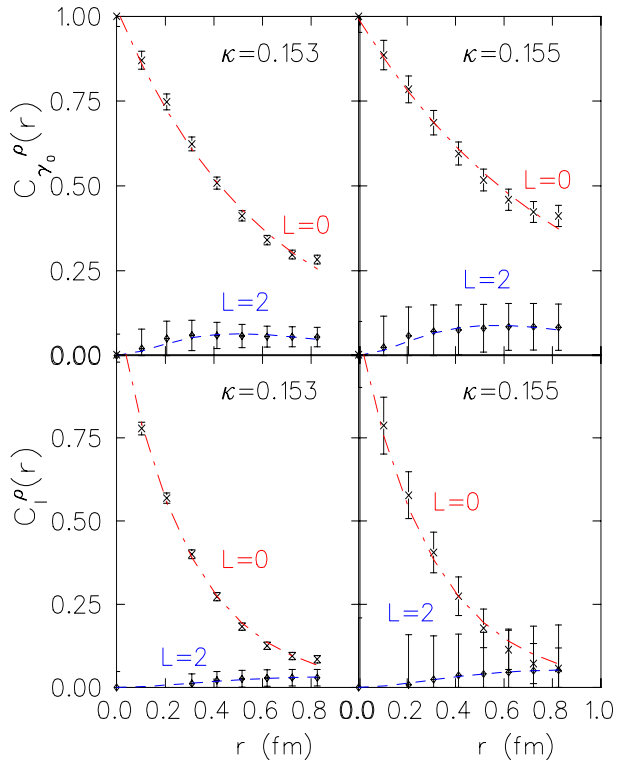}}
\caption{\label{fig:angular}Decomposition of the density-density 
correlator, $C_{\gamma_0}^{\rho}$, and  the matter density correlator,
$C_I^{\rho}$, for the rho meson into angular 
momentum part $L=0$ and $L=2$ at $\kappa=0.153$ (left) 
and $\kappa=0.155$ (right).}
\end{figure}

In Fig.~\ref{fig:rho3 asym} we show  the 
correlators for the spin-0 state of the rho, for quark separations
 along the spin axis ($z$)  and perpendicular to it ($x$).
We observe a $z-x$ asymmetry in the case of the 
charge distribution~\cite{AFT} whereas
 in the case of the matter distribution no statistically significant
asymmetry is present. 
The rho deformation  seen in Fig.~\ref{fig:rho3 asym}
can be made more quantitative by analysing 
the correlators  into a dominant $L=0$
and a suppressed $L=2$ state~\cite{Gupta2}:

\be <\rho_k({\bf 0})|\hat{j}_{\gamma_0}^u({\bf r})\hat{j}_{\gamma_0}^d({\bf 0})|\rho_k({\bf 0})>
= f_0(r) + \frac{(3x_k^2-r^2)}{3r^2} f_2(r)
\label{angular}
\ee

\noindent
where  $|\rho_k({\bf 0})>$ is a zero momentum state with polarization $k$.
The deformation, $\delta$, determined from the quadrupole moment, can be
defined as  
\be
\delta\equiv \frac{3}{4}\> \frac{\langle 3z^2-r^2 \rangle}{\langle r^2 \rangle} \quad.
\label{deformation}
\ee
Writing for  the symmetry axis $R_3 =\sqrt{\langle z^2 \rangle} = R_T+\epsilon$
where $R_T^2 = 1/2 \langle x^2+y^2 \rangle $ and assuming a small symmetry
deviation  $\epsilon$ we obtain
\be
\delta =\frac{\epsilon}{R_T} \sim \frac{R_3-R_T}{R}
\ee
where $R^2=1/3 \langle x^2+y^2+z^2 \rangle$ is the mean square radius.
Expanding the hadron wave function in $L=0$ and $L=2$ components as
$\psi({\bf r})= \alpha \phi_0(r) P_0(\cos\theta)
+ \beta\phi_2(r) P_2(\cos\theta)$, where $P_L$ are Legendre
Polynomials, we find for the deformation
\be
\delta=\frac{3}{5}\>\frac{\beta}{\alpha} \>
\frac{\langle \phi_0|r^2|\phi_2\rangle} {\langle \phi_0|r^2|\phi_0\rangle}
\ee
where we have neglected $\beta^2$ terms.
Taking $\phi_0(r) \sim \exp(-m_0 r) $ and
$\phi_2(r) \sim r^2 \exp(-m_2 r)$, and 
again neglecting $\beta^2$ terms
which appear when squaring the wave function to obtain the correlator
so that $f_0(r)=\alpha^2\phi_0^2(r)$ and 
$f_2(r)=3\alpha \beta\phi_0(r)\phi_2(r)$,
we fit  the resulting expression to the lattice data.
As shown in Fig.~\ref{fig:angular} for two $\kappa$ values, 
this ansatz provides a 
 good description of the rho charge density distribution.
 From the fits we find at the four $\kappa$ values that we have 
analysed a non-zero ratio $\beta/\alpha \sim 0.02$
 with an error of about 60\% which mainly arises from the
poor determination of the coefficient of the $L=2$ state.
This leads to a deformation of the order of $1\%$. The errors
in extracting $\delta$, within this approach, are large
since, in addition to the ratio $\beta/\alpha$,
this evaluation also involves  $m_0^5$ and $(m_0+m_2)^7$ amplifying further the error on
$\delta$.
A direct determination of the quadrupole
moment via Eq.~\ref{deformation}
 yields $\delta=0.03 \pm 0.01$, with a better control on the errors than
the value obtained from the angular decomposition.
Unquenching tends to increase this deformation, but the statistical
error also increases, so that no definite conclusion regarding
the importance of pion cloud contribution can be reached.
However, the fact that deformation is seen for the rho meson 
in the quenched theory casts doubt on  models which assume a spherical
core and attribute the deformation entirely  to the pion cloud~\cite{Buchmann}.
The same analysis for the matter density distribution
yields an $L=2$ component consistent with zero, as expected from the lack of $z-x$ asymmetry
demonstrated  in Fig.~\ref{fig:rho3 asym}. Also in the unquenched case, 
the matter
density asymmetry remains consistent with zero.

\begin{figure}[h]
\begin{center}
\epsfxsize=8.0truecm
\epsfysize=7.truecm
\mbox{\epsfbox{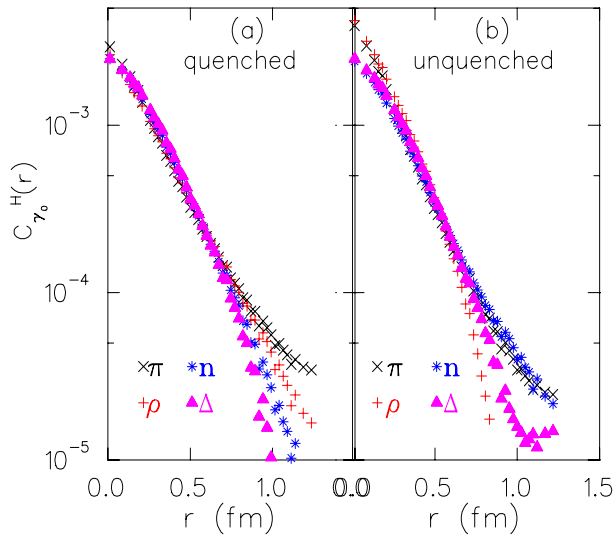}}
\caption{\label{fig:wfs_nf}
Comparison of the quenched matter density  distribution
 at $\kappa=0.153$ (left) to the  unquenched one at $\kappa=0.156$ (right)
for the pion (x's), the rho (crosses), the nucleon (stars) and the $\Delta^+$
(filled triangles)}
\end{center}
\end{figure}

\begin{figure}[h]
\begin{center}
\epsfxsize=8.0truecm
\epsfysize=7.truecm
\mbox{\epsfbox{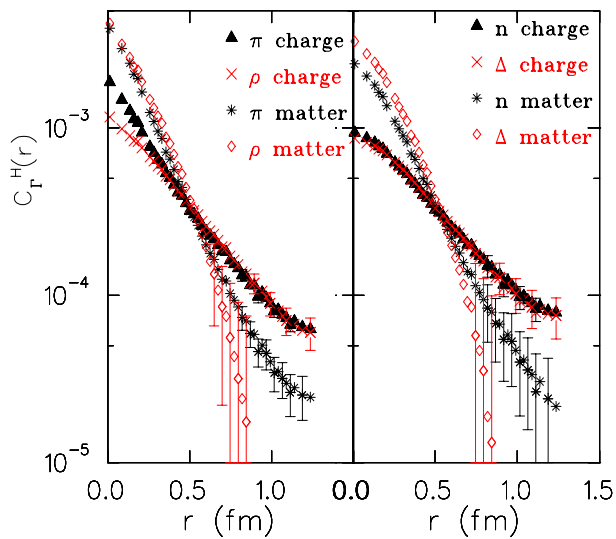}}
\caption{\label{fig:charge matter full}
Comparison of the charge density distribution to the matter distribution
for  the unquenched theory at $\kappa=0.156$. The notation is the
same as that of Fig.~\ref{fig:charge matter}.}
\end{center}
\end{figure}

The absence of a measurable 
deformation in the case of the rho matter distribution
has important implications for the mechanism that produces
the asymmetry in the rho charge distribution. Since the matter and charge operators
have the same  non-relativistic limit, 
this strongly suggests  that hadron
charge deformation is a relativistic effect.
That is particularly interesting
since it has strong implications for the validity of various models used 
in the study of nucleon deformation as well as in the evaluation of
transition matrix elements for $\gamma N\rightarrow\Delta$. In a number
of these models~\cite{Buchmann2} the quadrupole moments
are evaluated in the  non-relativistic constituent quark model
limit with D-state admixture or with two-body currents where 
one would expect the same results
for the charge and matter distributions.
In these models 
one thus expects that for  the rho meson
the charge and matter quadrupole moments are the same
up to an overall multiplicative factor  related to the different
electromagnetic and strong coupling constants.
The conclusion that the charge deformation is a relativistic
effect is confirmed by another  study employing
heavy quarks on a fine lattice. For these heavy
quark systems  no signal for charge deformation
was obtained~\cite{margarita}. If the pion cloud makes
a significant contribution to the charge deformation then
an analysis with lighter dynamical quarks should increase the
charge deformation
but it should also yield a deformation in the matter distribution.
Clearly, in view of such important phenomenological implications, it is
imperative to improve the lattice results, in particular
for the baryons, by using lighter quarks
on a larger volume
both in the quenched  and in the
unquenched theory, and determine
the amount of deformation in the chiral limit.

In order to investigate the
importance of dynamical quarks, we
analyse a subset of the  SESAM~\cite{SESAM} configurations.
The lattice spacing
determined from the rho mass in the chiral limit 
using the SESAM configurations
 is $a^{-1}= 2.3$~GeV~\cite{SESAM} which is  the same as for
the quenched theory at $\beta=6.0$, and therefore
the physical volume is the same as in our quenched calculation.
As we have already explained, we use this determination of the lattice spacing, which is applicable
in both the quenched and the unquenched theory,  to compare 
quenched and unquenched results.
For each of the the two values of the hopping parameter, 
$\kappa=0.156$ and $0.157$, we analysed 100 SESAM configurations.
The ratio  of the pion mass to rho mass is 0.83 at $\kappa=0.156$ and
0.76 at $\kappa=0.157$. These values are close to the quenched mass ratios
measured at $\kappa=0.153$ (0.84) and $\kappa=0.154$ (0.78) respectively,
allowing us to make pairwise quenched-unquenched comparisons.
Such comparison was made for the charge distribution in ref.~\cite{AFT}.
In Fig.~\ref{fig:wfs_nf} we compare the matter distribution for the quenched 
($\kappa=0.153$) and the unquenched ($\kappa=0.156$) theories. We observe
no unquenching effects for the baryons, whereas the matter density 
distribution for the  pion and rho increases
at the origin. Beyond about $0.7$~fm the statistical errors, 
not shown in the figure for clarity, become large
especially for the nucleon and the $\Delta^+$.
In Fig.~\ref{fig:charge matter full} we compare the unquenched charge and matter density distributions
at $\kappa=0.156$. Similar results with larger statistical errors are obtained for 
the lighter quark mass ($\kappa=0.157$),
indicating that the mass dependence of these
results is very weak.  As   
in the quenched case we observe a faster fall off of the matter density 
distribution
 as   compared to the charge density distribution.
Whereas unquenching 
increases the  rho asymmetry for the charge correlator  (see Fig.~17 of
ref.~\cite{AFT}),
it has no effect on the matter density distribution: Fig.~\ref{fig:rho3 asym}b
remains unchanged.

\section{Pseudoscalar density}

\begin{figure}[h]
\begin{center}
\epsfxsize=8.0truecm
\epsfysize=10.truecm
\mbox{\epsfbox{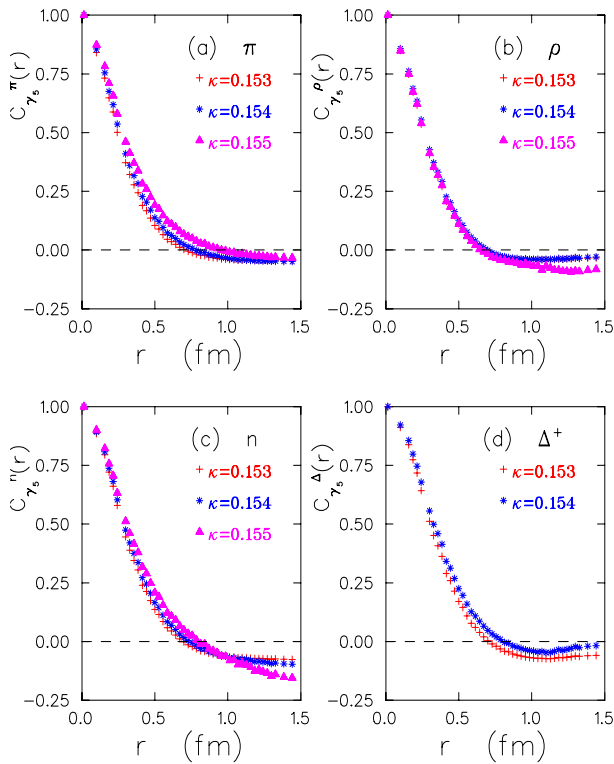}}
\caption{\label{fig:pseudoscalar}
The pseudoscalar density  at $\kappa=0.153$, $0.154$
and $0.155$ for (a) the pion, (b) the rho, (c) the nucleon and the (d) $\Delta^+$.}
\end{center}
\end{figure} 

\begin{figure}[h]
\begin{center}
\epsfxsize=8.0truecm
\epsfysize=11.truecm
\mbox{\epsfbox{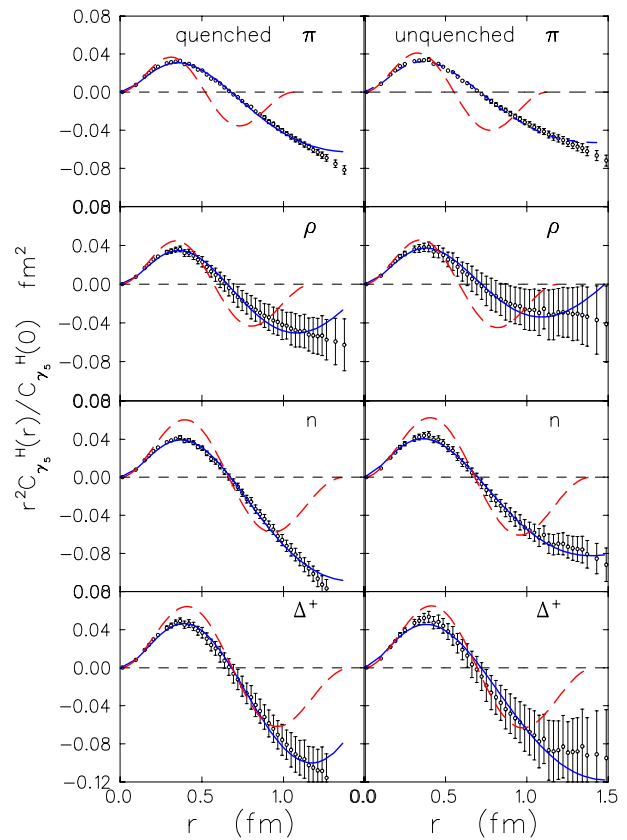}}
\caption{\label{fig:pseudoscalar k0153+k0156}The quenched (left) and
unquenched (right) pseudoscalar density at $\kappa=0.153$ and $\kappa=0.156$
respectively  with fits to 
$(a+bx^2+cx^4)e^{-m x}$ (solid line) and bag model results (dashed line).}   
\end{center}
\end{figure}

\begin{figure}[h]
\begin{center}
\epsfxsize=8.0truecm
\epsfysize=7.truecm
\mbox{\epsfbox{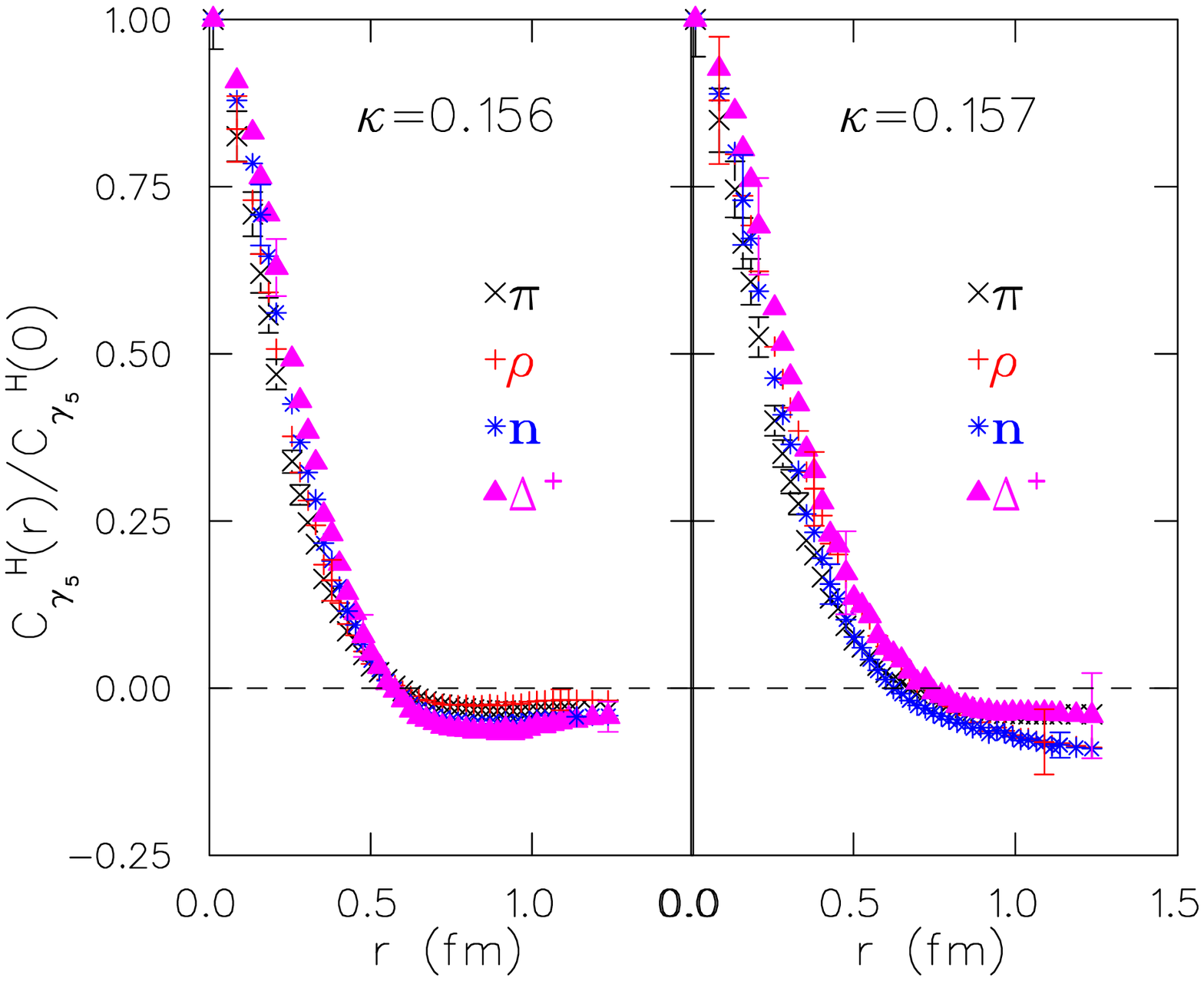}}
\caption{\label{fig:unquenched pseudoscalar}The unquenched 
 pseudoscalar density at $\kappa=0.156$ (left) 
and $\kappa=0.157$ 9right) for the pion (x's), the rho (crosses), the nucleon 
(stars) and the $\Delta^+$ (filled triangles).}
\end{center}
\end{figure}

The pseudoscalar density  is of theoretical interest,
especially because it can serve as an additional observable in our
comparison with bag model predictions. The lattice results
in the quenched theory for the pion, the rho, the nucleon and
the $\Delta^+$ are shown in Fig.~\ref{fig:pseudoscalar}
for $m_\pi/m_\rho= 0.84,\>0.78$ and $0.70$. Again we observe only a weak
dependence on the quark mass.
To compare with bag model predictions, which
will be discussed in the following section, it is best to  consider
the pseudoscalar density weighted with $r^2$. 
This
 is shown for both the quenched
and the unquenched theory at $\kappa=0.153$ and $\kappa=0.156$ respectively in Fig.~\ref{fig:pseudoscalar k0153+k0156}.
A reasonable fit is obtained using an exponential times a polynomial ansatz. 
We note here that in both the quenched and unquenched theory 
the long tail of the data and  
the integral of the fitted ansatz both
favor a non-zero integral. 
 In Fig.~\ref{fig:unquenched pseudoscalar} we show the unquenched pseudoscalar density for the four
hadrons under consideration normalized to unity at the origin.
We observe that the results are very similar for all hadrons,
with almost no
dependence on  the quark mass. 
As in the case of the matter distribution  
dynamical quark effects are small for these values of quark masses.
Before leaving this section we must stress that
these results on the pseudoscalar density are most useful in showing qualitative
features. Since we have used an unimproved Dirac operator which has order $a$ 
chiral symmetry violations, we expect sizable ultraviolet corrections.

\section{Comparison of lattice and  bag model results}

\begin{figure}[h]
\begin{center}
\vspace*{0.5cm}
\epsfxsize=8.0truecm
\epsfysize=10truecm
\mbox{\epsfbox{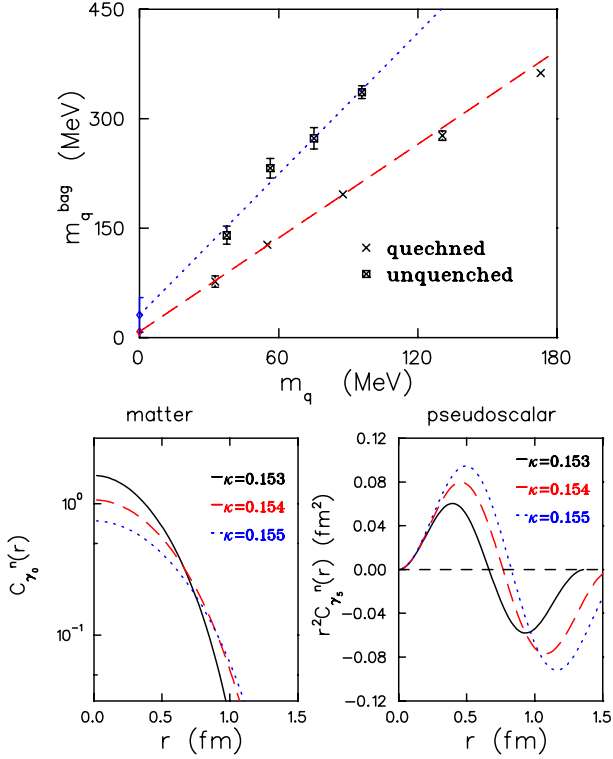}}
\caption{\label{fig:mqbag}Upper: The quark mass,
$m_q^{\rm bag}$, needed to obtain the lattice result for the nucleon mass
 versus the naive quark mass $m_q$. The  best straight line fits
to both the quenched (dashed) and unquenched (dotted) case are also displayed.
Lower: The  mass dependence of the matter (left) and pseudoscalar (right)
density distributions for the nucleon 
within the bag model with parameters fixed using procedure A
 at $\kappa=0.153, \> 0.154$ and $0.155$.}
\end{center}
\end{figure}

It is interesting to compare our lattice results to those
obtained in the bag model. We will consider only the lowest mode
of the free Dirac field in a spherical bag of radius $R$.
The radius $R$ is chosen so as to minimize the mass, $M(R)$, of the hadron
under consideration. There are four contributions to the
mass: the volume term $E_v=4\pi B R^3/3$, the zero point
energy, $E_0=-Z_0/R$, the kinetic energy of the quarks, $E_Q=n_q \omega$,
and the chromo-magnetic hyper-fine interaction energy 
calculated to
first order in perturbation theory which is proportional to $\alpha_{\rm bag}$~\cite{bag}. 
 $n_q$ is the number of quarks in the
hadron, $\alpha_{\rm bag}=g^2/4\pi$ where $g$ is the strong coupling constant,
and $\omega$ is the frequency of the lowest mode given explicitly below. 
We use two procedures to fix the bag model parameters $B, Z_0$ and $\alpha_{\rm bag}$:
In the first procedure, which we will refer to as procedure A,
we use the lattice values for the masses of the rho, the nucleon and the
$\Delta$ to fix the three parameters using as an input
 the naive quark mass, $m_q = 1/2a(1/\kappa-1/\kappa_c)$.
In Table~\ref{table:bag params} we give the bag parameters determined
from this procedure. In the quenched case, as the quark mass decreases
the bag model parameters nicely approach the parameters
determined using the experimental values of the
$\omega$ meson, the nucleon and the $\Delta$ and   $m_q=0$
also included in the last row of Table~\ref{table:bag params}.
For the unquenched case,  this trend is only clear for the
two smallest $\kappa$ values (heaviest quarks). For the larger $\kappa$ values
this is no longer clear and a possible explanation is
finite volume effects known to  become important at these small quark
masses.
 A second procedure suggested in ref.~\cite{Negele2}, referred to here as
procedure B,
is to use the bag model parameters as obtained in the last row of 
Table~\ref{table:bag params} and 
with these parameters fixed,   adjust the quark mass so that the lattice results for the
 mass of the nucleon is
reproduced at the given $\kappa$ value. Since we use the nucleon mass to define the quark
mass in this way, it is natural for this discussion to use the nucleon mass to set the
lattice scale $a$. This gives $a^{-1}=2.04(2)$~GeV for the quenched case and
$a^{-1}=1.88(7)$ for the unquenched theory. 
The quark mass obtained with procedure B is
denoted by $m_q^{\rm bag}$.
Both the bag model and the lattice phenomenology are consistent with
a linear dependence of the nucleon mass on the quark mass in the regime we explore.
Therefore we expect $m_q\propto m_q^{\rm bag}$.
In Fig.~\ref{fig:mqbag} we display the values obtained for 
 $m_q^{\rm bag}$ versus the naive quark
mass. 
The quenched data nicely fall on a straight line confirming
this expectation. We find that $m_q^{\rm bag}$ is  
about $2.1$ times the naive quark mass.
Extrapolating to
$m_q=0$ we obtain
$m_q^{\rm bag}=8 \pm 2$ MeV reasonably close to the expected zero value, especially
 since the  error does not include uncertainties of the order of 10\%
in setting the lattice scale.
The same analysis can also be done for the less accurate unquenched data. 
We find $m_q^{\rm bag}\sim 3 m_q$. Using
the SESAM results for the nucleon
mass at $\kappa=0.156,\> 0.1565, \> 0.157$ and
$0.1575$ and a  linear extrapolation
to the chiral limit, we obtain $m_q^{\rm bag}=31\pm 24$~MeV at $m_q=0$,
again
close to zero.
In the same figure we also show the bag model results for the
matter and pseudoscalar densities 
calculated with  procedure A  using the parameters 
extracted from fitting  the quenched data
 at $\kappa=0.153,\> 0.154$ and $\kappa=0.155$.
The bag model results show a stronger dependence on the quark mass in comparison
with the corresponding lattice results discussed in the previous sections.

\begin{table}
\caption{Bag model parameters extracted from fitting the masses of
the rho, the nucleon and the $\Delta$. The bag radii, $R_\pi$, $R_\rho$, $R_n$ 
and $R_\Delta$ for the pion, the rho, the nucleon and the $\Delta$ 
are given in GeV$^{-1}$ respectively.}
\begin{tabular}{|cccc||cccc|} \hline
\multicolumn{8}{|c|}{Quenched: $\beta=6.0$, $16^3\times 32$} \\ \hline
$m_q $ (GeV)  & $B^{1/4}$(GeV) & $Z_0$ &  $\alpha_{\rm bag}$  & $R_\pi$ & $R_\rho $ &  $R_{n}$  & $R_\Delta$ 
\\  \hline
0.174 & 0.209 & 2.274 & 0.169 & 2.71 & 2.99 & 3.48 & 3.59\\
0.131 & 0.182 & 2.245 & 0.196 & 3.09 & 3.47 & 4.00 & 4.15\\
0.088 & 0.169 & 2.228 & 0.239 & 3.26 & 3.76 & 4.29 & 4.49 \\ \hline
\multicolumn{8}{|c|}{$N_f=2$: $\beta=5.6$, $16^3\times 32$ } \\ \hline
$m_q$ (GeV)  & $B^{1/4}$(GeV) & $Z_0$ &  $\alpha_{\rm bag}$  & $R_\pi$ & $R_\rho $ &  $R_{n}$  & $R_\Delta$ 
\\  \hline
0.096 & 0.210 & 2.120 & 0.110 &  2.86 & 3.02 & 3.53 & 3.60 \\
0.076 & 0.195 & 2.096 & 0.136 &  3.06 & 3.28 & 3.80 & 3.89 \\
0.057 & 0.201 & 2.383 & 0.188 &  2.74 & 3.09 & 3.60 & 3.73 \\
0.038 & 0.171 & 2.227 & 0.188 &  3.33 & 3.70 & 4.28 & 4.43 \\
 \hline
\multicolumn{8}{|c|}{Continuum}\\ \hline
$m_q$ (GeV)  & $B^{1/4}$(GeV) & $Z_0$ &  $\alpha_{\rm bag}$  & $R_\pi$ & $R_\rho $ &  $R_{n}$  & $R_\Delta$ 
\\  \hline
0 & 0.146 &   1.86  & 0.55 & 3.33 & 4.67 & 4.97 & 5.44 \\ \hline
\end{tabular}
\label{table:bag params}
\end{table}

In order to evaluate the  various density distributions we
expand the quark fields in terms of bag eigenmodes keeping
only the lowest mode~\cite{Negele2}:
\beq
q({\bf r}) &=& \sum_{s,c}\biggl(b_s^c\psi_s({\bf r}) + d_s^{c\>\dagger}\phi_s({\bf r}) \biggr) \nonumber \\
\psi_s &=&   {i\> f(r) U_s \choose - g(r) {\bf{\sigma}} .\hat{{\bf r}} U_s } 
\hspace*{1cm} \phi_s({\bf{r}}) = C\gamma_0 \psi^*_s({\bf r})\nonumber \\
f(r)& = & \frac{N(x)}{\sqrt{4\pi}} 
\biggl (\frac{\omega+m}{\omega} \biggr )^{1/2} j_0(xr/R)\nonumber \\
g(r)&=&\frac{N(x)}{\sqrt{4\pi}}\biggl (\frac{\omega-m}{\omega} \biggr )^{1/2} j_1(xr/R)
\label{bag eigemodes}
\eeq
where $U_s$ are two component Pauli spinors, $C=\gamma_2\gamma_0$ is the charge
operator, $j_i$
are spherical Bessel functions, $m$ is the value of the quark mass
which here is  either $m_q$ or $m_q^{\rm bag}$
 and $N(x)$ is fixed by normalizing the eigenmode in the bag.
The frequency, $\omega(m,R)$, of the lowest mode 
is given by $\omega=1/R\;(x^2+m ^2 R^2)^{1/2}$ where $x(mR)$ is the solution of 
the eigenvalue equation $(1-mR-\omega R) \tan x = x$. The
superscript $c$ on the quark annihilation operators $b_s^c$ and on the
anti-quark creation operators  $d_s^{c\>\dagger}$
denotes flavour and colour quantum numbers.
In terms of the lowest eigenmode the density operators are
given by
\beq
\hat{\rho}_{\gamma_0}({\bf r}) &=& \sum_{s,c}\left(b_s^{c \> \dagger} b_s^{c}
-d_s^{c \> \dagger} d_s^{c} \right)\left(f(r)^2+g(r)^2\right) \\
\hat{\rho}_I({\bf r}) &=& \sum_{s,c}\left(b_s^{c \> \dagger} b_s^{c}
+d_s^{c \> \dagger} d_s^{c} \right)\left(f(r)^2-g(r)^2\right) \\
\hat{\rho}_{\gamma_5}({\bf r}) &=& -2 i\sum_{s,s',c}
\left(b_{s}^{c \> \dagger} b_{s'}^{c}U^\dagger_s {\bf \sigma}.\hat{\bf r}U_{s'} \right . \nonumber \\
&\>& \left . \hspace*{1cm} -d_s^{c \> \dagger} d_{s'}^{c}U^\dagger_s {\bf \sigma}^*.\hat{\bf r}U_{s'} \right) f(r)g(r)
\label{rho}
\eeq
and  the charge, matter and pseudoscalar  correlators 
by the following expressions:
\be
\left \langle 
\begin{array}{ccc} \pi  |&  & |\pi \\
    \rho | &\hat{j}_{\gamma_0}^u({\bf r})\>  \hat{j}_{\gamma_0}^d({\bf r}') & |\rho \\
     n  | &  & | n \\
 \Delta| &  &|\Delta 
\end{array} \right \rangle 
= C \left (\begin{array}{c} -1 \\ -1\\ 1\\1\end{array}\right )
\> \begin{array}{c}\left(f(r)^2+g(r)^2 \right). \\ 
\left( f(r')^2+g(r')^2 \right)
\end{array}
\label{bag_charge}
\ee

\be
\left \langle 
\begin{array}{ccc}  \pi  |&  & |\pi \\
\rho |  &\hat{j}_{I}^u({\bf r})\>  \hat{j}_{I}^d({\bf r}')  & |\rho 
\\  n  | &  & | n \\
 \Delta| &  &|\Delta 
\end{array} \right \rangle 
=  C' \left ( \begin{array}{c} 1 \\1 \\1\\ 1\end{array} \right )
 \begin{array}{c} \left( f(r)^2-g(r)^2 \right).\\
 \left( f(r')^2-g(r')^2 \right)
\end{array}
\label{bag_matter}
\ee
                                
\be
\left \langle 
\begin{array}{ccc}  \pi  |&  & |\pi \\
\rho | &\hat{j}_{\gamma_5}^u({\bf r})\>  \hat{j}_{\gamma_5}^d({\bf r}')  & |\rho \\
  n  | &  & | n \\
 \Delta| &  &|\Delta 
\end{array} \right \rangle 
= C'' \left ( \begin{array}{c} 1\\-1/3\\ 2/3\\-2/3\end{array}\right )
\> \begin{array}{c} 
4 \hat{\bf r}.\hat{\bf r}' f(r)g(r) \\
f(r')g(r')\end{array}
\label{bag_pseudoscalar}
\ee
with constants $C,\> C'$ and $C''$ independent of the hadron state.
The minus sign for mesons in the charge correlator
arises because in the density operator the term involving anti-quarks
comes with a negative sign. This term is positive in the case
of the matter density giving the same sign for mesons and baryons. 
This shows  explicitly that the difference between charge and matter 
density distributions is due to the opposite sign of the lower  components
of the Dirac spinor which it is a relativistic effect.

In the case of the pseudoscalar correlator given by 
Eq.~\ref{bag_pseudoscalar} the results were obtained by averaging
over spin projections of the physical states i.e. 
for the rho we averaged over the states
with $J_z=\pm 1$ and $J_z=0$. 
In order to understand the signs for the pseudoscalar density consider
the pion where the spin of the quark and the anti-quark are opposite.
Since $|\pi\rangle=1/\sqrt{2}\sum_s(b_{+}^{u \> \dagger} d_{-}^{d \dagger}
-b_{-}^{u \> \dagger} d_{+}^{d \>\dagger})|0\rangle$, we have from
Eq.~\ref{rho} a contribution of the form $\pm {\bf \sigma}.\hat{\bf r}U_{\pm}
U_{\mp}{\bf \sigma}^*.\hat{\bf r'}U_{-}$ yielding the result $-\hat{\bf r}.\hat{\bf r}'$
and thus an overall positive sign.
The rho state with $J_z=0$ is orthogonal to the pion resulting in the opposite
sign. Similar considerations lead to the opposite sign between the nucleon
and the $\Delta^+$.

\begin{figure}[h]
\begin{center}
\epsfxsize=8.0truecm
\epsfysize=11.truecm
\mbox{\epsfbox{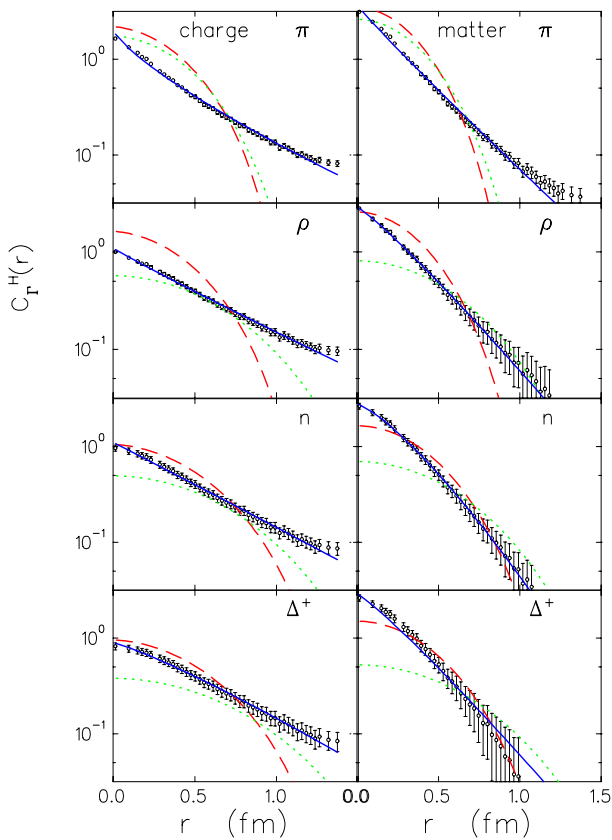}}
\caption{\label{fig:wfs_bag}
Comparison of quenched charge (left) and matter distribution (right)
 at $\kappa=0.153$ with bag model results. The dashed line is obtained
by using procedure A as described in the text and the dotted line
 by using  procedure B.
The solid line is a fit to $a\exp(-m r^{\alpha})$.}
\end{center}
\end{figure}

In Fig.~\ref{fig:wfs_bag} we compare the lattice results for the charge
and matter density distributions with the bag model results from Eqs.~\ref{bag_charge}
and \ref{bag_matter}. The dashed lines show the results obtained
with procedure A.
The dotted lines are the bag model results obtained within
procedure B.
 Both procedures  fail
to reproduce the correct radial dependence of the charge
distribution. The results obtained using  procedure A
provide an overall better description in the case
of the baryon matter distribution.
To make the comparison quantitative, we evaluate
the root mean square radius, which provides a measure
of the width of the distributions. In the bag model one usually computes 
the expectation value of ${\bf r}^2$ for the lowest eigenmode.
In our previous study of the charge density distribution
~\cite{AFT} we used the quark model definition
for the charge radius which
for mesons is given by
\beq
<r^2_{\rm ch}> &=&  \sum_q e_q \langle({\bf r}_q- {\bf R}_{\rm cm})^2\rangle
 \nonumber \\
&=& \frac{\sum_q e_q\int d^3r\> ({\bf r}/2)^2\> C_{\gamma_0}({\bf r})}{ \int d^3r\>C_{\gamma_0}({\bf r})}
\label{charge rms}
\eeq
where ${\bf R}_{\rm cm}$ is the coordinate of the
 center of mass and $e_q$ is the electric charge of the quarks.
A corresponding definition for baryons can only be used if one
knows the charge density distribution in terms of the two
relative coordinates which requires the evaluation of three current
correlation functions. Here we only evaluate the one-density
baryon charge distribution and therefore the quark model
 definition can not be applied. 
For simplicity and for direct comparison with the the bag model 
radial width we calculate for both mesons and baryons 
\be
\langle r^2\rangle = \frac{1}{2}\frac{\int d^3r\> {\bf r}^2\> C_{\Gamma}({\bf r})}{ \int d^3r\>C_{\Gamma }({\bf r})} \quad,
\label{matter rms} 
\ee
where for $\Gamma=\gamma_0$ we obtain the charge mean square  radius,  $r^2_{ch}$,
and for  $\Gamma=I$ the matter mean square  radius,  $r^2_{m}$.
For degenerate quarks and for the case where the meson wave function
is a product of single particle radial wave functions like
in the bag-model, the factor of $1/2$ corrects for summing over 
the charge root mean square radius of each quark  as it is done 
 in Eq.~\ref{matter rms} in using the relative quark distance square.

The lattice results for the charge and matter  radii are collected
in Table~\ref{table:lattice rms}.
The main observation is that hadron sizes show very little quark mass
dependence in the quenched case, at least for the range of quark masses
considered, but increase under the effect of dynamical quarks.
One should keep in mind, however, that finite volume effects on these
values have not been investigated here, but could be significant
at the smaller quark masses.

Despite the failure of the bag model in describing the individual radial shape
of the distributions 
it produces reasonable results for the relative  widths of
 the charge to the matter distribution as can be seen from
Table~\ref{tab:table1}. Both lattice and bag model
consistently predict
a broader charge than matter distribution,
with a very weak mass dependence.
In Table~\ref{tab:table1} we also give the bag-model results for the
charge radius which show a stronger dependence on the quark
mass as well as on the channel than the lattice results.

In Fig.~\ref{fig:wfs_bag}, in addition to the
bag-model results, we have also included a fit of the lattice
 data to the
ansatz $\exp(-mr^{\alpha})$.
As it can be seen this simple ansatz 
 provides the best description to the data.
In fact,  except for the pion, a good fit 
to all other 
charge and matter correlators is obtained by taking $\alpha=1$,
which gives a $\chi^2$ per degree
of freedom $<1$. In this case we can identify asymptotically
 the mass $m$ of the pure
exponential to the mass of a propagating meson. 
To fit the matter distributions 
we find that we need
a  mass of 
approximately twice 
that required to fit the charge distributions of these
 hadrons. 
The asymptotic behaviour of the density correlators can be analysed
using a tree graph approximation  based on the fact
that at large distances light hadrons  dominate~\cite{Negele}.
In our work
 the ratio of  the pion to the rho mass is always 
larger than one-half. In this case, the analysis of ref.~\cite{Negele} shows
that the density correlators should decay exponentially  with the
mass of the rho meson, expect for the rho correlator whose 
exponential decay  at very
large distances is governed by the pion mass.
A similar analysis for the matter correlator predicts asymptotically
an exponential decay determined from the lightest scalar in the theory.
In the continuum the lightest scalar ($f_0$) is heavier than the rho by
200~MeV and one thus expects asymptotically
a faster fall off of the matter correlator as compared to the charge
correlator.
Unfortunately, the values we obtain for the masses, $m$, of the exponential fall off
from the fits  can not be identified with the mesons
of the theory.
In general we find  {\em slower} exponential decay than expected
from the mass of the lightest
meson which should dominate the asymptotic behaviour.
Presumably, our lattice  is not
large enough to probe the asymptotic behaviour of the correlators.
The pion
correlators require $\alpha<1$, or the sum of two exponentials governing the
short and long distance decays respectively.
If we nevertheless use the single exponential ansatz
with $\alpha=1$ also for this channel the ratio of mass values needed to fit
the charge and matter density distributions 
is about 1.4 instead of two as for the other hadrons decreasing
slightly with the quark mass. 
This value of $\sim 1.4$
 is thus in accord with the value of $\sim 1.6$
 found from the study of heavy light mesons
in ref.~\cite{Green}.   

\begin{table}
\caption{\label{table:lattice rms} Square root of the
 charge ($r^2_{ch}$) and matter ( $r^2_m$)  radii
in fm extracted from the lattice data. We  use the nucleon mass to convert
to physical units.}
\begin{ruledtabular}
\begin{tabular}{|c||lc|lc|lc|} 
\multicolumn{7}{|c|}{Quenched: $\beta=6.0$, $16^3\times 32$} \\ \hline
 $\kappa$&\multicolumn{2}{c|}{$0.153$}&\multicolumn{2}{c|}{0.154} 
 &\multicolumn{2}{c|}{0.155}\\
 & $\sqrt{r^2_{ch}}$  & $\sqrt{r^2_m}$  & $\sqrt{r^2_{ch}}$ & $\sqrt{r^2_m}$ & $\sqrt{r^2_{ch}}$ &$\sqrt{r^2_m}$  \\ \hline
 $\pi$      & 0.444(4) & 0.396(6)  & 0.454(4) & 0.417(6) & 0.465(8)& 0.440(7)\\ 
 $\rho$     & 0.457(5) & 0.397(18) & 0.470(6) & 0.404(37)& 0.482(11) & - \\
 $n$        & 0.457(4) & 0.383(16) & 0.465(6) & 0.384(31)  & 0.473(11) & - \\
 $\Delta^+$ & 0.464(6) & 0.390(19) & 0.469(9) & 0.360(76)  & 0.447(77) & -\\
\hline
\multicolumn{7}{|c|}{$N_f=2$: $\beta=5.6$, $16^3\times 32$ } \\ \hline
 $\kappa$&\multicolumn{2}{c|}{$0.156$}&\multicolumn{2}{c|}{0.157} & &
 \\
 & $\sqrt{r^2_{ch}}$  & $\sqrt{r^2_m}$  & $\sqrt{r^2_{ch}}$ & $\sqrt{r^2_m}$ & &  \\ \hline
 $\pi$      & 0.471(3) & 0.411(7)  & 0.490(3) & 0.439(7)  & & \\ 
 $\rho$     & 0.481(6) & 0.370(52) & 0.510(6) & 0.373(121)& & \\
 $n$        & 0.485(5) & 0.427(14) & 0.509(7) & 0.414(57)  &  &  \\
 $\Delta^+$ & 0.489(9) & 0.384(55) & 0.522(9) & 0.436(102)  &  & 
\end{tabular}
\end{ruledtabular}
\end{table}

\begin{table} 
\caption{\label{tab:table1} Ratio of the 
expectation value of ${\bf r}^2$ for the charge 
distribution over that for the  matter distribution in
quenched lattice QCD and bag model using procedure A. For $\kappa=0.155$ we give
the ratio only for the pion since for the other hadrons
the matter radius was too noisy. The root mean square  charge 
radius in fm in the bag-model using procedure A is also given.}
\begin{ruledtabular}
\begin{tabular}{|c||lc|lc|lc|} 
\multicolumn{7}{|c|} {$<{ r}_{ch}^2>/<{r}_m^2>$}  \\ \hline
 $\kappa$&\multicolumn{2}{c|}{$0.153$}&\multicolumn{2}{c|}{0.154} 
 &\multicolumn{2}{c|}{0.155}\\
 & lattice & bag & lattice & bag & lattice & bag \\ \hline
 $\pi$ & 1.24(3) & 1.35 & 1.19(2) & 1.36  & 1.10(2)& 1.37\\ 
 $\rho$ &1.33(7) & 1.35 & 1.35(17)  & 1.36&- &  1.37\\
 $n$ &1.40(8) &1.34 & 1.45(16) & 1.35 &- & 1.36 \\
 $\Delta$ &1.41(9) & 1.34 & 1.7(5) & 1.35 & - & 1.36 \\ \hline 
\multicolumn{7}{|c|} {$\sqrt{r_{ch}^2}$ in the bag model (fm)} \\ \hline
 $\kappa$&\multicolumn{2}{c|}{$0.153$}&\multicolumn{2}{c|}{0.154} 
 &\multicolumn{2}{c|}{0.155}\\ \hline
 $\pi$ & \multicolumn{2}{c|}{0.383} & \multicolumn{2}{c|}{0.439}& \multicolumn{2}{c|}{0.466} \\ 
 $\rho$ &\multicolumn{2}{c|}{0.422} & \multicolumn{2}{c|}{0.492}& \multicolumn{2}{c|}{0.537} \\ 
 $n$ & \multicolumn{2}{c|}{0.488} & \multicolumn{2}{c|}{0.565}& \multicolumn{2}{c|}{0.612} \\ 
 $\Delta$ & \multicolumn{2}{c|}{0.503} & \multicolumn{2}{c|}{0.586}& \multicolumn{2}{c|}{0.639} 
\end{tabular}
\end{ruledtabular}
\end{table}

In the case of the pseudoscalar density 
the bag model predicts that the integral $\int d^3r \;  C_{\gamma_5}^H(r)$ is
 zero. This prediction of the bag model is clearly 
seen   in Fig.~\ref{fig:mqbag}.  It was for this
reason that  in Fig.~\ref{fig:pseudoscalar k0153+k0156} we chose to show
the lattice quenched and unquenched results for the
 pseudoscalar density weighted by $r^2$.
As can be seen from Fig.~\ref{fig:pseudoscalar k0153+k0156}, for 
both the quenched and the unquenched results,
 the bag model prediction is worse for the pion, since the lattice correlator 
turns negative at a larger distance. 
This is not surprising since the bag model is known to be worse for the pion.
For the baryons
on the other hand the node of the bag model distribution
coincides with that of the lattice data. In
all cases  the lattice data have a long tail which is not reproduced
in the bag model, and which favors a non-zero integral.  
 However a careful thermodynamic and continuum extrapolation,
on larger and finer lattices,
is required for this quantity, especially with Wilson fermions as used here.
For the smaller quark masses noise at the 
tail of the distribution makes the situation even worse.

\section{Conclusions}
To check our lattice procedure we performed an
analysis of the charge density distributions using
various types of sources. For a lattice 
of temporal extent $32a$ considered here and within
our statistics we found  that local and Wuppertal smeared sources
give consistent results when we insert the 
current operator at time separation
from the source of $t=8a$, which is the maximal allowed
separation given the anti-periodic boundary conditions
in the temporal direction. The results are also in agreement with
those obtained using Dirichlet boundary conditions which
allow larger time separations. Therefore, for the lattice parameters
used in  this study, local sources are a suitable choice
  for the evaluation of density 
distributions since they have less gauge
noise than smeared ones and
 the temporal extent is large enough to make 
contributions from high excited states negligible.
Wall sources are shown  to filter
the ground state less effectively and  at the
maximal time separation of $8a$ yield results that are not
in  agreement with the rest.
Using the lattice conserved current, which has a different $a$-dependence
as compared to the continuum current,   these correlators 
show
finite lattice spacing effects 
near the origin. 

Our  main conclusion from the comparison of quenched and
unquenched results for the  charge, the matter and the
pseudoscalar densities is that no sizable 
 unquenching effects are observed for quark masses in the range of 300-100~MeV.
The
charge density distribution is, in all cases,
broader than the  matter density. This is what is expected if a tree level
classical approximation is used to describe
  the asymptotic behaviour of these correlators.
However the masses of the propagating lightest mesons which
determine the exponential fall off can not
be identified with the mesons of the theory. To observe the
true asymptotic behaviour larger lattices are needed. 
For the pion, the lattice results are in agreement with 
the experimentally extracted ratio of the charge to the matter radius
of $1.15(9)$~\cite{Chou}.
For baryons, the lattice indicates a charge  radius about
20\% larger than the  matter radius. 
This effect is well reproduced by the bag model.
However the bag model does a poor job in the
description of the  radial dependence, especially in the case
of the mesons. Instead, the charge and the
matter distributions are well described by the simple
ansatz  $\exp(-m r^\alpha)$, with $\alpha=1$ except for the pion.
One prediction of the bag model is that the 
 the volume integral
of the pseudoscalar
density  is zero. Lattice data for both the  quenched and the unquenched
theory 
favor a negative value. 
A larger lattice is required to settle this issue.

The deformation seen in the rho charge distribution is absent in the matter
distribution, both in the quenched and the unquenched theory. This 
observation suggests a relativistic origin for the deformation. 
This important issue deserves a more extended study, with lighter quarks
and larger volumes.

\vspace*{0.5cm}

\noindent
\underline{Acknowledgments:}
The $SU(3)$ $16^3\times 32$ quenched
lattice configurations were obtained from the
Gauge Connection archive~\cite{NERSC}. We thank the SESAM collaboration
for giving us access to their  dynamical lattice configurations.
A. T. acknowledges the Levendis Foundation.

\end{document}